\documentclass[letterpaper,journal]{IEEEtran}

\usepackage[T1]{fontenc}
\usepackage[utf8]{inputenc}
\usepackage{cite}
\usepackage{url}
\usepackage[final]{microtype}

\usepackage{graphicx}
\usepackage{amsmath,amssymb}
\usepackage{tikz}
\usetikzlibrary{arrows.meta,calc,shapes.geometric}
\usepackage{booktabs}
\usepackage{array}
\usepackage{tabularx}
\usepackage[table]{xcolor}
\newcommand{\stagecontent}[2]{%
  \begin{minipage}[c][2.52cm][t]{4.05cm}
    \centering
    \hspace*{0.70cm}%
    \parbox[c][0.72cm][c]{3.28cm}{%
      \centering
      \fontsize{7.4}{8.2}\selectfont
      #1%
    }\par
    \vspace{0.2mm}
    \includegraphics[
      width=3.92cm,
      trim=0 125 0 125,
      clip
    ]{#2}%
    \vfill
  \end{minipage}%
}

\tikzset{
  stage/.style={
    rounded corners=3pt,
    line width=0.8pt,
    inner sep=1pt,
    align=center,
    anchor=center
  },
  stage1/.style={
    stage,
    draw=blue!75!black,
    fill=blue!3
  },
  stage2/.style={
    stage,
    draw=cyan!55!black,
    fill=cyan!3
  },
  stage3/.style={
    stage,
    draw=violet!70!black,
    fill=violet!3
  },
  stage4/.style={
    stage,
    draw=orange!80!black,
    fill=orange!3
  },
  stage5/.style={
    stage,
    draw=green!55!black,
    fill=green!3
  },
  gate/.style={
    diamond,
    aspect=1.0,
    draw=black!75,
    fill=white,
    line width=0.7pt,
    minimum size=5.0mm,
    inner sep=0pt
  },
  flow/.style={
    -{Latex[length=2.0mm,width=1.35mm]},
    draw=black!75,
    line width=0.75pt
  },
  revise/.style={
    -{Latex[length=1.8mm,width=1.2mm]},
    dashed,
    draw=black!75,
    line width=0.65pt
  },
  flowlabel/.style={
    font=\fontsize{6}{6.8}\selectfont,
    text=black,
    fill=white,
    inner sep=0.6pt
  }
}

\newcommand{\agentloop}[2]{%
  \begin{scope}[
    shift={($(#1.north west)+(0.40,-0.43)$)}
  ]
    \draw[
      #2,
      line width=0.65pt,
      -{Latex[length=1.35mm,width=0.9mm]}
    ]
      (-150:0.24)
      arc[start angle=-150,end angle=25,radius=0.24];

    \draw[
      #2,
      line width=0.65pt,
      -{Latex[length=1.35mm,width=0.9mm]}
    ]
      (30:0.24)
      arc[start angle=30,end angle=205,radius=0.24];

    \draw[black!75,line width=0.55pt]
      (-0.090,-0.090) rectangle (0.090,0.090)
      (-0.052,-0.052) rectangle (0.052,0.052);

    \foreach \p in {-0.060,0,0.060}{
      \draw[black!75,line width=0.45pt]
        (\p,0.090)--(\p,0.135)
        (\p,-0.090)--(\p,-0.135)
        (0.090,\p)--(0.135,\p)
        (-0.090,\p)--(-0.135,\p);
    }
  \end{scope}%
}

\newcommand{\humanat}[1]{%
  \fill[black!72]
    ($(#1.center)+(0,0.77)$)
    circle[radius=0.085];

  \path[
    draw=black!72,
    fill=black!65,
    line width=0.45pt,
    rounded corners=1pt
  ]
    ($(#1.center)+(-0.14,0.61)$)
    rectangle
    ($(#1.center)+(0.14,0.44)$);

  \draw[black!72,line width=0.65pt]
    ($(#1.center)+(0,0.44)$)
    --
    (#1.north);
}

\newcommand{\horizontalgate}[3]{%
  \node[gate] (#2)
    at ($(#1.east)!0.5!(#3.west)$) {};

  \humanat{#2}

  \draw[flow]
    (#1.east) -- (#2.west);

  \draw[flow]
    (#2.east) -- (#3.west);

  \coordinate (#2ret)
    at ($(#1.east)+(0,-0.87)$);

  \coordinate (#2elbow)
    at (#2.south |- #2ret);

  \draw[revise]
    (#2.south)
    --
    (#2elbow)
    --
    (#2ret);

  \node[flowlabel]
    at ($(#2.east)!0.5!(#3.west)+(-0.12,0.36)$)
    {$\checkmark$ Pass};

  \node[flowlabel,anchor=north]
  at ($(#2elbow)!0.5!(#2ret)+(0.15,-0.12)$)
  {$\times$ Revise};
}

\title{
  Exploiting LLM Agents for Trustworthy AutoResearch
  in Wireless Communications
}
\author{
Yuan Guo$^{\star}$,~\IEEEmembership{Member,~IEEE},
Zixiang Ren$^{\star}$,~\IEEEmembership{Member,~IEEE},
Jie Xu,~\IEEEmembership{Fellow,~IEEE},
Liang Hong, \\
Fan Liu,~\IEEEmembership{Senior Member,~IEEE},
and Rui Zhang,~\IEEEmembership{Fellow,~IEEE}%
\thanks{$^{\star}$Yuan Guo and Zixiang Ren contribute equally to this work.}%
\thanks{Yuan Guo and Jie Xu are with the School of Science and Engineering and the Future Network of Intelligence Institute, The Chinese University of Hong Kong (Shenzhen), Shenzhen, China (e-mail: guoyuan@cuhk.edu.cn; xujie@cuhk.edu.cn). Jie Xu is the corresponding author.}%
\thanks{Zixiang Ren and Rui Zhang are with the Department of Electrical and Computer Engineering, National University of Singapore, Singapore (e-mail: zixiang\_ren@nus.edu.sg; elezhang@nus.edu.sg).}%
\thanks{Liang Hong is with Sun Yat-Sen University, Guangzhou, China (e-mail: hongliang@sysu.edu.cn).}%
\thanks{Fan Liu is with the National Mobile Communications Research Laboratory, Southeast University, Nanjing, China (e-mail: fan.liu@seu.edu.cn).}%
}
\begin{document}
\maketitle

\begin{abstract}
Large language model (LLM) agent-enabled AutoResearch is attracting growing interest across scientific disciplines, in which LLMs are leveraged for knowledge synthesis, multistep planning, code generation, tool invocation, and iterative refinement, thus automating the research lifecycle, from hypothesis generation and experimentation to analysis and manuscript preparation. Wireless communications is particularly suitable for this paradigm. This is due to the fact that advances in this field often rely on fundamental-limit analysis, system optimization, and protocol design, all supported by mature mathematical, simulation, and optimization toolchains, as well as standards, measurement, and digital-twin platforms. However, integrating autonomous agents into rigorous wireless research workflows requires traceable processes and verifiable evidence. This article presents a general trustworthy wireless AutoResearch framework. This framework employs typed research contracts, version-controlled artifacts, independent validators, and bounded agent authority to connect hypothesis generation with system modeling, mathematical formulation, algorithm design, code generation, simulation, reproducible claims, and manuscript preparation. We conduct a case study on integrated sensing and communication (ISAC) for unmanned aerial vehicles (UAVs). It is shown that the proposed framework can generate innovative research ideas and, with appropriate expert intervention, produce a manuscript whose evaluation score is comparable to or higher than those of related IEEE conference and letter papers. These findings demonstrate the effectiveness of LLM agents in orchestrating authoritative scientific tools and versioned research artifacts, while underscoring the importance of human–AI collaboration.
\end{abstract}

\section{INTRODUCTION}

Large language model (LLM) agents combine LLMs with agent harnesses to move beyond single-shot text generation, transforming them into autonomous systems capable of multi-step problem solving \cite{yao2023react,zhong2026harness}. While LLMs provide flexible reasoning and language-based coordination capabilities, agent harnesses provide the runtime infrastructure required for persistent and controllable execution, including tool interfaces, context management, permissions, resource budgets, provenance tracking, validation mechanisms, and stopping rules. This model–harness integration enables LLM agents to decompose complex objectives into manageable subtasks and coordinate their execution, thereby supporting applications across software engineering, data analysis, robotics, network operations, and scientific discovery \cite{zhong2026harness}.

Among these application domains, scientific research is particularly well suited to agentic automation because scientific discovery is inherently iterative and typically spans multiple stages, including literature review and gap identification, hypothesis generation, experimentation and interpretation, manuscript preparation, and peer review. Many of these stages can be supported by LLM reasoning coupled with external scientific tools. Building on this potential, prior systems such as AI Scientist \cite{lu2026aiscientist}, AutoResearchClaw \cite{liu2026autoresearchclaw}, and Robin with laboratory-in-the-loop experimentation \cite{ghareeb2026robin} have demonstrated that LLM agents can already coordinate substantial portions of the research lifecycle across diverse scientific fields.

Wireless communications, particularly physical-layer (PHY) research, provides a natural domain for this paradigm. A typical PHY research workflow spans literature review and idea generation, system modeling and assumptions, performance and fundamental-limit analysis, problem formulation, algorithm design, simulation and code implementation, numerical evaluation, and conclusions supported by evidence. These processes benefit from a mature ecosystem of scientific tools that can be orchestrated by LLM agents. For instance, LLM agents can coordinate web search and knowledge synthesis for literature review and idea generation; Sionna supports programmable link-level simulation and differentiable ray tracing \cite{hoydis2022sionna,hoydis2023sionnart}; and MATLAB, CVX, and related numerical tools provide complementary capabilities for analysis and optimization. Together, these tools provide executable support across the PHY research lifecycle, enabling LLM agents to test and iteratively refine their decisions.

Despite this potential, ensuring the trustworthiness of LLM-agent-enabled wireless AutoResearch remains challenging. Here, trustworthiness means that scientific claims are traceable to explicit assumptions, versioned artifacts, and verifiable evidence. Accordingly, such systems should be assessed not by the fluency of their outputs but by their ability to conduct executable studies, support independent validation and reproducibility, and appropriately bound conclusions. This process-oriented perspective requires preserving the structure of physical models, maintaining consistency and dependencies across research artifacts, separating generation from evaluation, documenting failed attempts, and ensuring that claims do not exceed the supporting analytical, simulation, or experimental evidence.

In this article, we propose a general framework for wireless AutoResearch, building upon our prior Wireless AutoResearch Agent (WARA) developed for automating wireless optimization research \cite{guo2026wara}. Going beyond the specific optimization scenarios considered in WARA, the proposed framework covers the complete research lifecycle through five major stages: literature review and idea generation; system modeling and fundamental-limit analysis; problem formulation and system optimization; simulation, experimentation, and result generation; and scientific insight synthesis and manuscript preparation. The framework employs {\it research contracts} to preserve physical and mathematical meaning, {\it versioned artifacts} to expose dependencies, {\it independent validators} to evaluate intermediate outputs, and {\it bounded agent authority} to maintain human control over consequential scientific decisions. It is worth noting that the five stages can either be integrated into an iterative end-to-end workflow, or executed independently as agent-assisted tasks. Accordingly, the framework supports varying levels of research autonomy: humans may approve each stage transition, or intervene only at designated stages, or permit bounded autonomous iteration across all stages within a predefined research contract.

Building on this framework, we present a representative PHY case study on integrated sensing and communication (ISAC) for UAVs \cite{liu2022isac,zeng2019accessing}. The case study covers the five research stages, with experts examining the generated artifacts and guiding their refinement.\footnote{Supporting research artifacts are available at
	\url{https://github.com/guoyuan-dotcom/TrustworthyWirelessAutoResearch}.}  The agents first generate candidate ideas with verifiable novelty and execution feasibility. Based on evaluations by dedicated agents and human experts, we select the idea ``Near-Field ISAC for Six-Dimensional UAV Pose Estimation'' and develop it from system modeling through manuscript preparation. With appropriate expert intervention, the resulting manuscript receives evaluation scores comparable to or higher than those of related IEEE conference and letter papers. These results demonstrate the effectiveness of LLM agents in orchestrating authoritative scientific tools and versioned research artifacts, while underscoring the importance of human–AI collaboration.

It is worth emphasizing that the proposed framework is different from prior applications of LLM agents in wireless communications. In the literature, ComAgent employs multiple agents to formulate and solve predefined wireless optimization tasks \cite{li2026comagent}. The AI Telco Engineer establishes an executable generate--evaluate--select loop for predefined PHY/MAC problems and objective functions \cite{aoudia2026telcoengineer}. In addition, recent work on agentic autonomous networks translates service intents into configuration, optimization, fault management, and control actions \cite{demirel2026intents}. These approaches demonstrate the value of domain grounding, tool invocation, executable feedback, and multi-agent coordination. However, they generally operate within a predefined problem formulation, objective function, action space, or operational envelope. In contrast, the proposed framework targets the broader scientific discovery process, including formulating and refining research questions, developing models and algorithms, generating and validating evidence, and establishing which scientific claims are justified by the available evidence.

\section{Why Wireless AutoResearch: From Scientific Challenges to Accelerated Discovery}
This section discusses why wireless communications is a natural domain for AutoResearch and highlights its benefits in accelerating scientific discovery, expanding innovation spaces, enhancing system evaluation, and promoting reproducible research.
\vspace{-3mm}
\subsection{Wireless Communications as a Natural Testbed for AutoResearch}

The emergence of wireless AutoResearch is driven by the convergence of two trends: the growing complexity of wireless research and the rapid advancement of LLM agents. Modern wireless innovation increasingly requires the integration of theories, models, algorithms, simulations, and experiments, while LLM agents can facilitate this process through knowledge synthesis, hypothesis generation, code execution, tool orchestration, and iterative refinement. This potential is particularly significant because of three distinctive characteristics of the field: cross-domain combinatorial innovation, structured mathematical discovery, and executable research loops supported by mature simulation ecosystems.

\subsubsection{Cross-Domain Knowledge Fusion for Combinatorial Innovation}A defining characteristic of modern wireless innovation is the increasing importance of cross-domain integration. Many emerging research directions arise from combining wireless communications with other disciplines, such as electromagnetics, antenna technologies, sensing, control, computing, and machine learning. For example, ISAC integrates communication and sensing theories \cite{liu2022isac}; movable antennas combine electromagnetic propagation with spatial optimization \cite{zhu2024movable}; near-field communications exploit electromagnetic field characteristics beyond conventional far-field models \cite{Zeng10496996}; and wireless AI integrates communication systems with machine learning methodologies \cite{li2026comagent}.

Such innovations are fundamentally combinatorial: new research opportunities often emerge from discovering previously unexplored connections among existing concepts, models, and techniques. However, exploring this rapidly expanding design space is increasingly challenging for individual researchers due to the explosive growth of scientific literature and the complexity of interdisciplinary knowledge. LLM agents provide a new approach by combining large-scale literature retrieval, knowledge graphs, and domain-specific reasoning. An agent can explore candidate combinations across different fields, identify relevant theories and methodologies, retrieve related prior work, and generate research hypotheses beyond the scope of individual researchers. 

Importantly, a meaningful scientific idea requires more than combining concepts. It must establish compatible assumptions, identify a causal or mathematical mechanism, distinguish itself from the nearest prior art, and provide experiments capable of validating the proposed explanation. These tasks require structured reasoning and evidence-based validation, which can be supported by well-designed LLM agents.

\subsubsection{Structured Mathematical Discovery: From Fundamental Limits to System Optimization}Wireless communications has a strong mathematical foundation, making many research problems naturally compatible with agent-assisted scientific discovery. Two representative categories are fundamental-limit analysis and system optimization.

Fundamental-limit analysis aims to characterize the ultimate performance achievable under a specific operational model and a set of assumptions. Wireless communications has developed rich theoretical foundations, including information theory, stochastic modeling, estimation theory, and optimization theory, providing extensive mathematical knowledge that can be leveraged by LLM agents. An agent can retrieve relevant theories, identify hidden assumptions, decompose complex conjectures into intermediate lemmas, and utilize symbolic algebra systems, proof assistants, optimization tools, or counter-example searches to assist theoretical development. However, numerical agreement alone cannot establish theoretical validity. A trustworthy AutoResearch system must preserve the information structure of the problem, maintain mathematical assumptions and regularity conditions, track dependencies among intermediate results, and ultimately rely on human-validated proofs for theoretical claims.

System optimization represents another major class of wireless research problems well suited for agentic automation. Wireless optimization typically involves translating system scenarios into mathematical variables, objectives, constraints, and implementable algorithms. LLM agents can assist in problem formulation, identify optimization structures, select appropriate solvers, generate baseline algorithms, implement simulations, and explore alternative designs. However, the scientific validity of optimization results depends not only on achieving improved numerical performance, but also on ensuring fair comparisons under consistent assumptions and practical constraints on energy, latency, and computational resources.

\subsubsection{Executable Research Loops Enabled by Wireless Simulation Ecosystems}Another important advantage of wireless research is the availability of mature modeling and evaluation platforms. Wireless systems can often be represented through mathematical models and evaluated through executable simulations before practical deployment. Modern platforms such as Sionna provide programmable link-level, channel, antenna, and ray-tracing simulations, while MATLAB, CVX, and network simulators offer complementary capabilities for modeling, optimization, and validation. Together, these tools establish executable research loops in which LLM agents can iteratively design algorithms, generate code, conduct simulations, analyze results, identify failures, and refine solutions. Such closed-loop interactions enable agents to test hypotheses, evaluate designs, and continuously improve research outcomes.
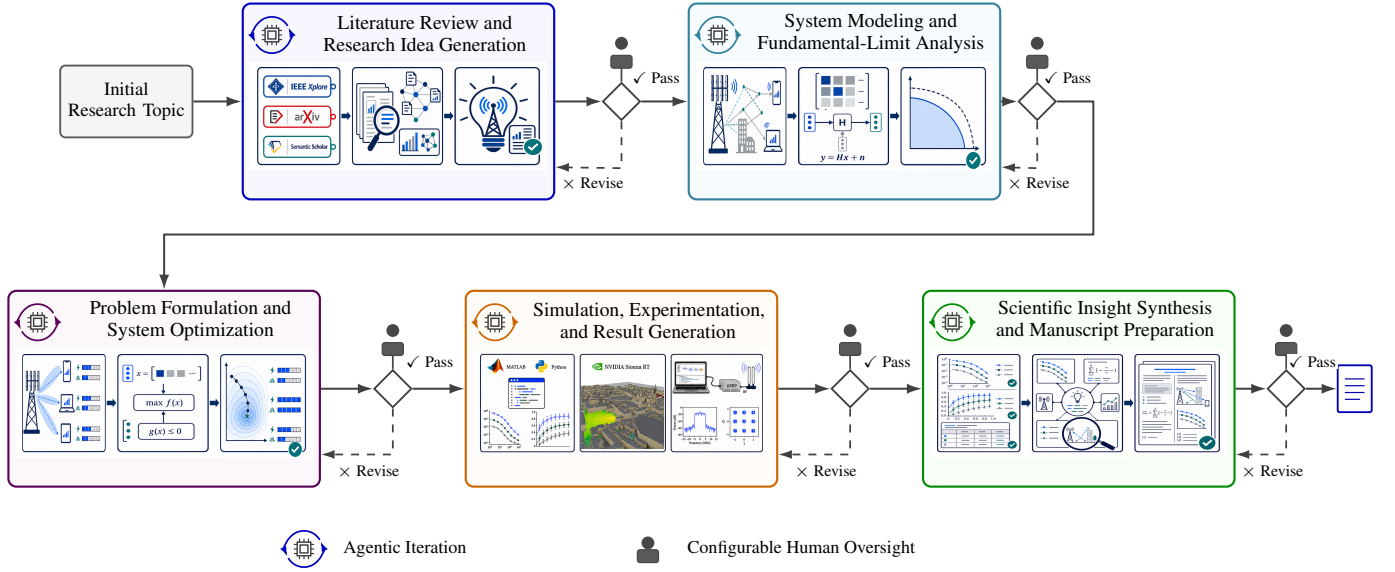
\begin{figure*}[!t]
\centering

\makebox[\textwidth][c]{%
\begin{tikzpicture}


\node[
  draw=blue!70!black,
  fill=blue!3,
  draw=black!65,
  fill=black!4,
  rounded corners=2pt,
  line width=0.7pt,
  minimum width=1.75cm,
  minimum height=0.95cm,
  inner sep=2pt,
  align=center,
  font=\fontsize{7.0}{7.8}\selectfont
] (topic) at (-6.55,3.80) {
  Initial\\Research Topic
};

\node[stage1] (f1) at (-2.95,3.80) {%
  \stagecontent
    {Literature Review and\\Research Idea Generation}
    {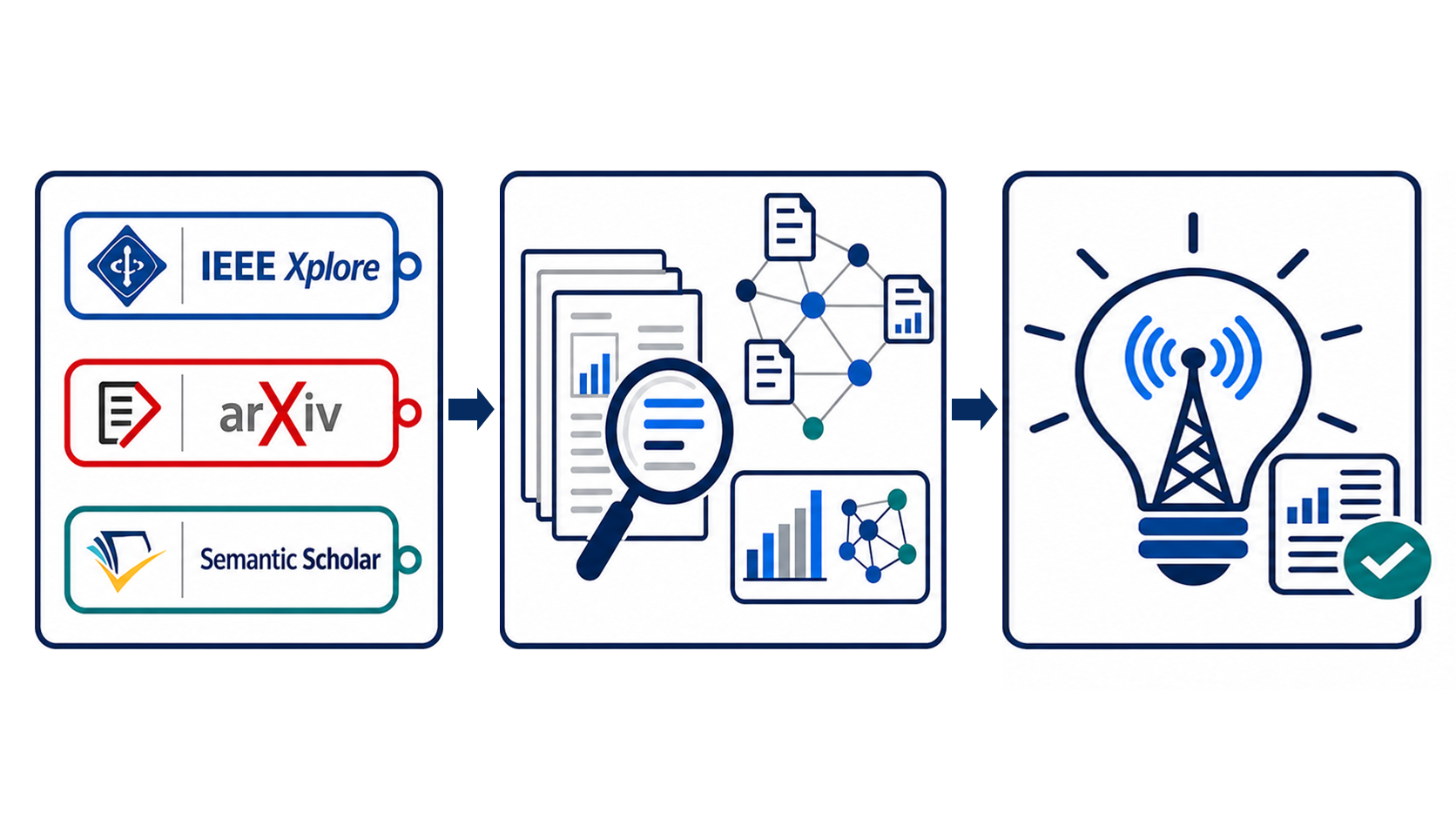}%
};

\draw[flow]
  (topic.east) -- (f1.west);

\node[stage2] (f2) at (2.95,3.80) {%
  \stagecontent
    {System Modeling and\\Fundamental-Limit Analysis}
    {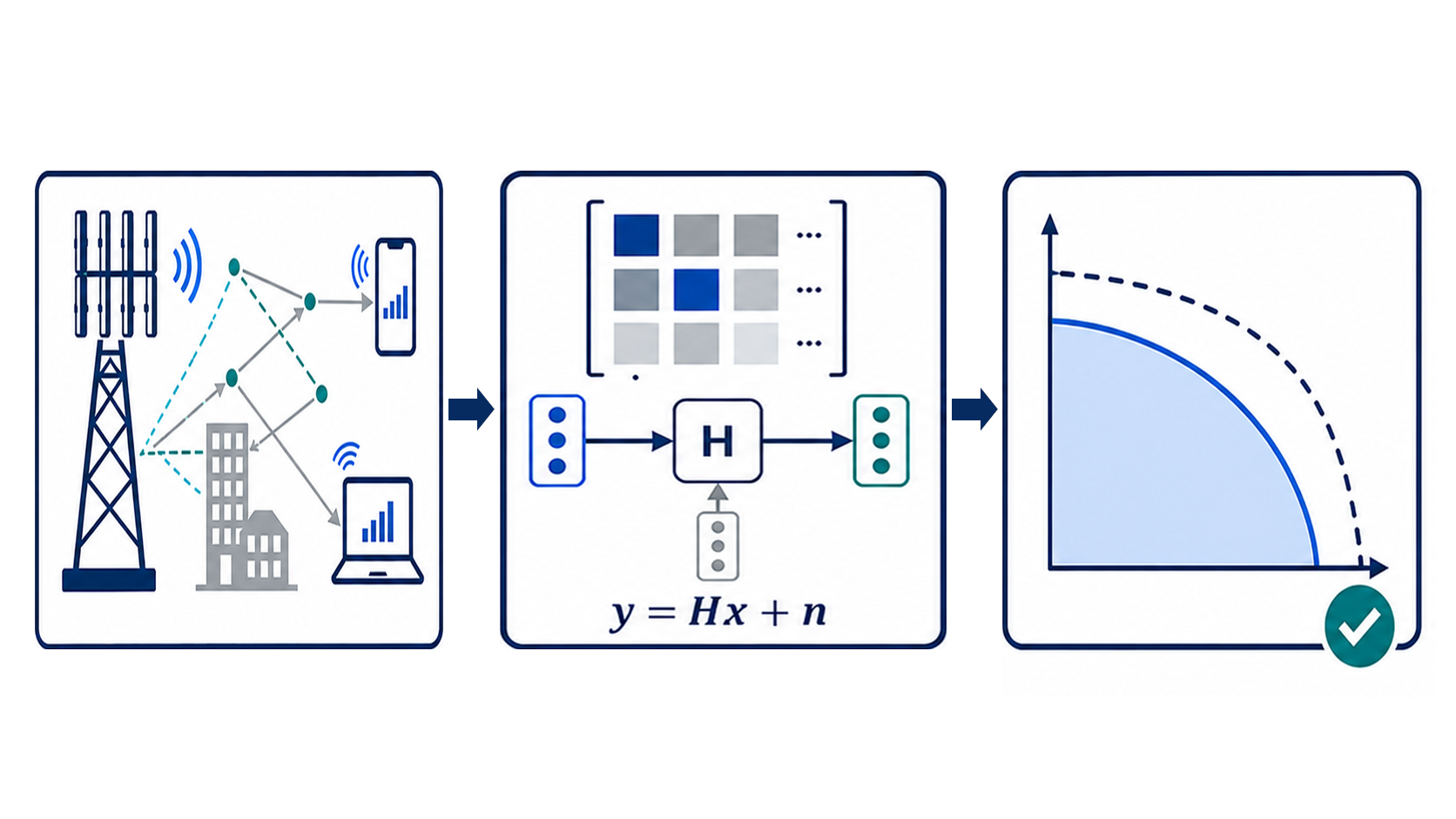}%
};


\node[stage3] (f3) at (-6.05,0) {%
  \stagecontent
    {Problem Formulation and\\System Optimization}
    {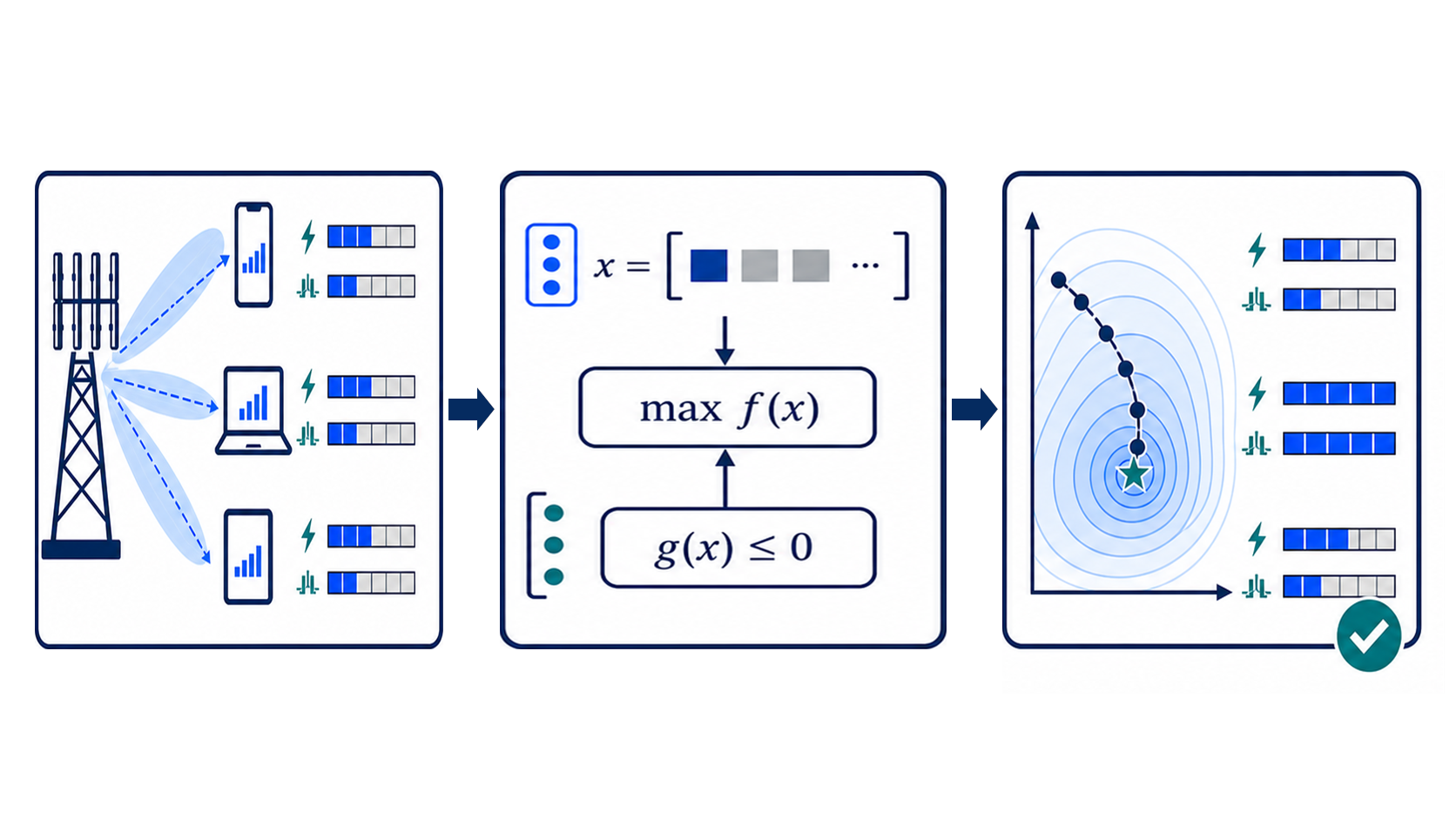}%
};

\node[stage4] (f4) at (0,0) {%
  \stagecontent
    {Simulation, Experimentation,\\and Result Generation}
    {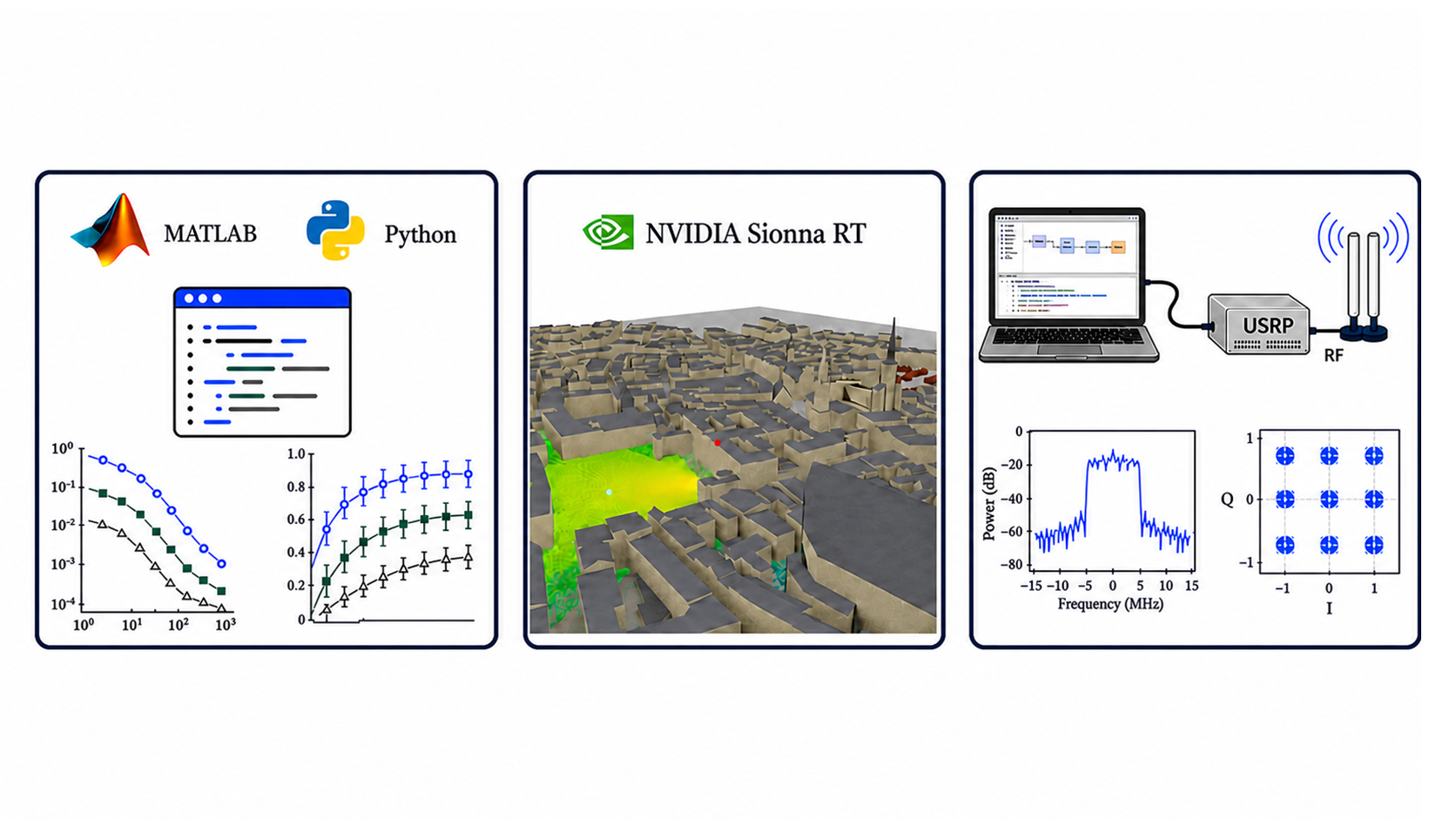}%
};

\node[stage5] (f5) at (6.05,0) {%
  \stagecontent
    {Scientific Insight Synthesis\\and Manuscript Preparation}
    {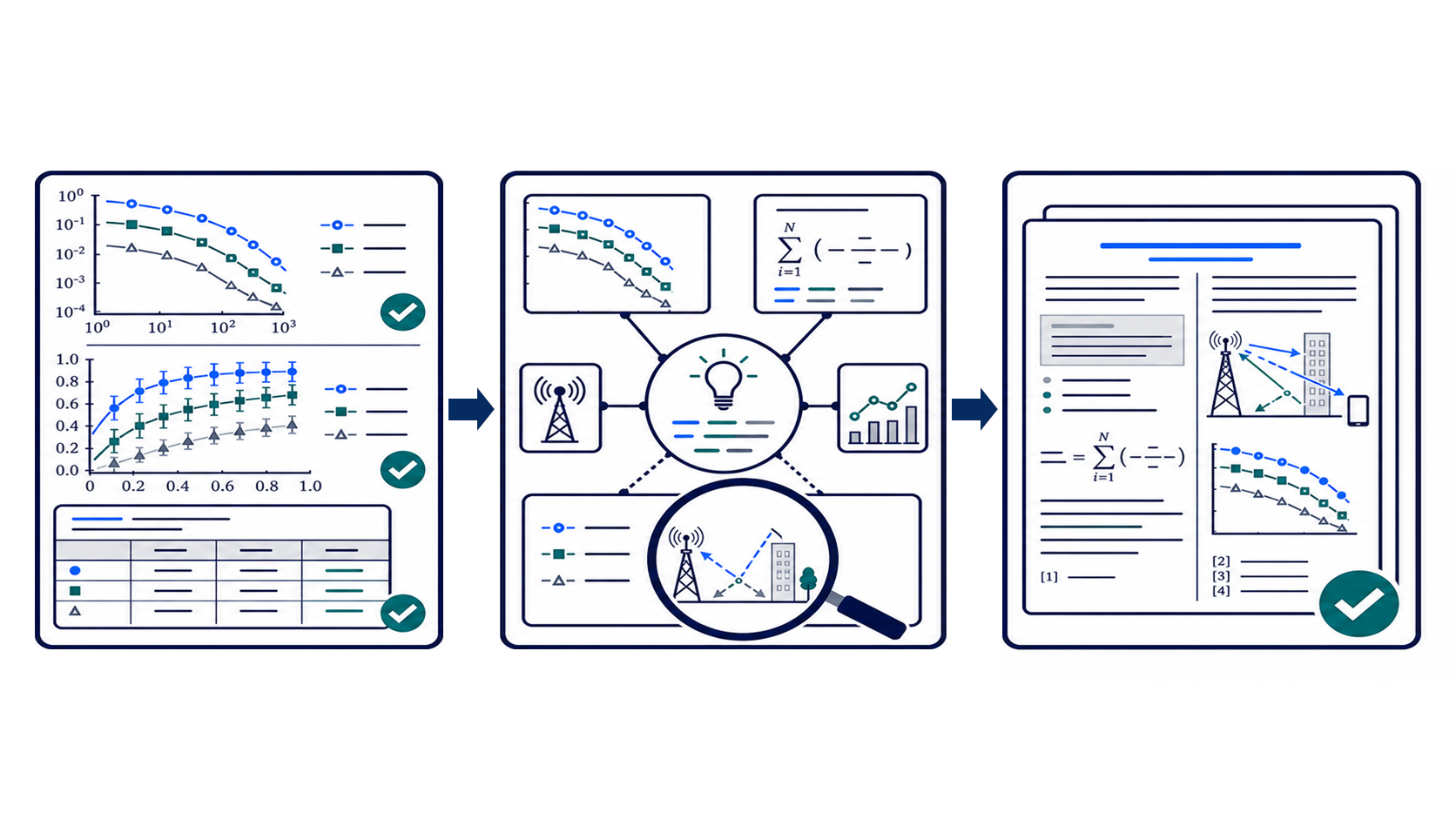}%
};


\agentloop{f1}{blue!75!black}
\agentloop{f2}{cyan!55!black}
\agentloop{f3}{violet!70!black}
\agentloop{f4}{orange!80!black}
\agentloop{f5}{green!55!black}


\node[gate] (g1) at (0,3.80) {};
\node[gate] (g2) at (5.50,3.80) {};

\humanat{g1}
\humanat{g2}


\draw[flow]
  (f1.east) -- (g1.west);

\draw[flow]
  (g1.east) -- (f2.west);

\draw[revise]
  (g1.south)
  --
  ++(0,-0.62)
  --
  ($(f1.east)+(0,-0.87)$);

\node[flowlabel]
  at ($(g1.center)!0.52!(f2.west)+(0,0.31)$)
  {$\checkmark$ Pass};

\node[flowlabel,anchor=west]
  at ($(f1.east)+(0.05,-1.09)$)
  {$\times$ Revise};


\draw[flow]
  (f2.east) -- (g2.west);

\draw[flow]
  (g2.east)
  --
  ++(0.48,0)
  --
  ++(0,-1.76)
  -|
  (f3.north);

\draw[revise]
  (g2.south)
  --
  ++(0,-0.62)
  --
  ($(f2.east)+(0,-0.87)$);

\node[flowlabel]
  at ($(g2.center)+(0.40,0.31)$)
  {$\checkmark$ Pass};

\node[flowlabel,anchor=west]
  at ($(f2.east)+(0.05,-1.09)$)
  {$\times$ Revise};


\node[
  draw=blue!70!black,
  fill=white,
  rounded corners=0.7pt,
  line width=0.7pt,
  minimum width=4.5mm,
  minimum height=6.2mm,
  inner sep=0pt
] (paper) at (9.70,0) {};

\draw[blue!70!black,line width=0.45pt]
  ($(paper.center)+(-0.13,0.13)$)
  --
  ($(paper.center)+(0.13,0.13)$)

  ($(paper.center)+(-0.13,0.04)$)
  --
  ($(paper.center)+(0.13,0.04)$)

  ($(paper.center)+(-0.13,-0.05)$)
  --
  ($(paper.center)+(0.13,-0.05)$)

  ($(paper.center)+(-0.13,-0.14)$)
  --
  ($(paper.center)+(0.07,-0.14)$);


\horizontalgate{f3}{g3}{f4}
\horizontalgate{f4}{g4}{f5}
\horizontalgate{f5}{g5}{paper}


\begin{scope}[shift={(-4.20,-2.12)}]
  \draw[
    blue!75!black,
    line width=0.65pt,
    -{Latex[length=1.35mm,width=0.9mm]}
  ]
    (-150:0.24)
    arc[start angle=-150,end angle=25,radius=0.24];

  \draw[
    blue!75!black,
    line width=0.65pt,
    -{Latex[length=1.35mm,width=0.9mm]}
  ]
    (30:0.24)
    arc[start angle=30,end angle=205,radius=0.24];

  \draw[black!75,line width=0.55pt]
    (-0.090,-0.090) rectangle (0.090,0.090)
    (-0.052,-0.052) rectangle (0.052,0.052);

  \foreach \p in {-0.060,0,0.060}{
    \draw[black!75,line width=0.45pt]
      (\p,0.090)--(\p,0.135)
      (\p,-0.090)--(\p,-0.135)
      (0.090,\p)--(0.135,\p)
      (-0.090,\p)--(-0.135,\p);
  }
\end{scope}

\node[
  anchor=west,
  font=\fontsize{6.8}{7.4}\selectfont
] at (-3.80,-2.12) {
  Agentic Iteration
};

\fill[black!72]
  (0.35,-2.04)
  circle[radius=0.085];

\path[
  draw=black!72,
  fill=black!65,
  line width=0.45pt,
  rounded corners=1pt
]
  (0.21,-2.13)
  rectangle
  (0.49,-2.30);

\node[
  anchor=west,
  font=\fontsize{6.8}{7.4}\selectfont
] at (0.75,-2.12) {
  Configurable Human Oversight
};

\end{tikzpicture}%
}
	\vspace{-7mm}
\caption{
  Workflow for trustworthy wireless AutoResearch.
}
	\vspace{-5mm}
\label{fig:wireless-research-lifecycle}

\end{figure*}
\vspace{-3mm}
\subsection{Transformative Benefits of Wireless AutoResearch}

\subsubsection{Accelerating Knowledge Discovery and Research Productivity}

Wireless research requires substantial effort in literature review, mathematical analysis, simulation development, and performance evaluation, particularly due to its increasingly cross-disciplinary nature. LLM agents can accelerate these activities by reviewing papers and comparing technical approaches in parallel. Such parallelized knowledge processing enables researchers to explore broader knowledge and devote more effort to scientific reasoning and innovation.

\subsubsection{Expanding Design Spaces through Automated Algorithm Exploration}

Wireless research typically involves iterative cycles of mathematical formulation, algorithm development, implementation, and evaluation. LLM agents can shorten this cycle by assisting these stages. Moreover, agents can systematically explore alternative architectures, algorithms, and parameter configurations, enabling researchers to investigate \textcolor{black}{a much larger design space than conventional manual approaches}.

\subsubsection{Enabling Large-Scale and Realistic System Evaluation}

Traditional wireless evaluations are often constrained by human effort and implementation complexity, resulting in simplified channel models, limited simulation scenarios, or manually selected test cases. By orchestrating advanced simulation tools, LLM agents can configure ray-tracing simulations and evaluate designs across realistic propagation environments, mobility patterns, and network conditions, thereby moving beyond narrow benchmark scenarios.

\subsubsection{Improving Reproducibility and Scientific Transparency}

Reproducibility is essential for trustworthy scientific research. LLM agents can reduce the effort required to reproduce wireless results by organizing research artifacts, documenting assumptions, and recreating experimental workflows. By maintaining explicit connections among assumptions, models, algorithms, code, and results, AutoResearch systems can promote more transparent and reproducible wireless research. This capability is particularly valuable in wireless communications, where subtle differences in channel assumptions, information availability, simulation settings, or evaluation metrics can significantly affect reported conclusions.
\vspace{-3mm}
\subsection{Challenges Toward Trustworthy Wireless AutoResearch}

Despite these significant opportunities, trustworthy wireless AutoResearch remains challenging. Scientific discovery requires more than generating plausible ideas, equations, or simulation results. A valid research contribution must rest on consistent assumptions and a clear mechanism, differ from prior work, and support its claims with evidence. Theoretical claims require valid assumptions and proof dependencies; optimization results require fair comparisons under consistent information and resource constraints; and simulations must align models, implementations, parameters, and metrics. These challenges motivate the framework and validation mechanisms developed in the following section.

\section{Evidence-Centered Framework for Trustworthy Wireless AutoResearch}

This section presents a trustworthy wireless AutoResearch framework in which scientific claims are traceable to the underlying research questions, modeling assumptions, analytical foundations, and supporting evidence. As shown in Fig.~\ref{fig:wireless-research-lifecycle}, the proposed framework consists of five stages covering the entire wireless research lifecycle, from an initial wireless topic to evidence-supported scientific findings. Throughout this lifecycle, LLM agents leverage wireless-domain knowledge, task-specific reasoning, and specialized tools to generate and iteratively refine research outputs. Contract-governed validation and configurable human oversight regulate how these outputs advance through the workflow.
\vspace{-3mm}
\subsection{Research Stages of Wireless AutoResearch}

We decompose wireless AutoResearch into five stages, each equipped with one or more agents to perform specialized research activities and generate corresponding outputs, as elaborated below.

\subsubsection{Literature Review and Research Idea Generation}

Given an initial wireless research topic, LLM agents formulate focused queries and use scholarly platforms, such as IEEE Xplore and arXiv, alongside document-analysis tools to retrieve and synthesize relevant literature and standards. Parallel retrieval and cross-document analysis allow them to cover a larger evidence base more rapidly than manual review while retaining source traceability. By comparing how prior studies formulate and evaluate the target wireless problem, the agents identify technical gaps and develop candidate ideas grounded in plausible physical or mathematical mechanisms. The candidate ideas are evaluated through multi-agent debate against explicit criteria, including novelty, technical feasibility, and testability. Each agent assigns criterion-level scores and provides supporting rationales, after which the scores and critiques are aggregated to rank the candidate ideas~\cite{liu2026autoresearchclaw,guo2026wara}. Subject to the configured human-check policy, a high-potential candidate idea is selected and refined into a well-defined and testable research hypothesis for subsequent system modeling and fundamental-limit analysis.

\subsubsection{System Modeling and Fundamental-Limit Analysis}

In this stage, LLM agents establish a wireless system model for the selected hypothesis and analyze the mechanisms that determine its achievable performance. The model specifies the network configuration, channel and signal relationships, information available to each entity, resource constraints, and performance metrics, with every assumption assigned a clear physical meaning and validity range \cite{guo2026wara}. Based on the established system model, the agents quantify the effects of system parameters and physical-layer conditions on the relevant performance metrics, and derive performance bounds or fundamental tradeoffs when appropriate. The analysis reveals the dominant design factors and feasible operating region, providing a principled basis for defining the decision variables, objective, and constraints in the subsequent optimization stage.

\subsubsection{Problem Formulation and System Optimization}

The focus of this stage now shifts from characterizing achievable performance to determining how the wireless system should be designed under its physical and operational constraints. LLM agents identify the controllable degrees of freedom and formulate an optimization problem that captures the performance objective, information available to network entities, and resource limitations. They first assess whether the formulation is convex and directly tractable. If it is nonconvex, the agents identify the source of nonconvexity and determine whether the problem admits an equivalent reformulation or requires an iterative, search-based, or learning-based solution. Based on the selected solution route, they develop an implementable algorithm with explicit computational steps and stopping conditions, and establish its convergence, complexity, or optimality properties when supported. Consistency checks verify that the algorithm addresses the original formulation, while detected mismatches trigger targeted revision \cite{aoudia2026telcoengineer,guo2026wara}. The resulting formulation and algorithm specification define the technical input for subsequent simulation and experimentation.

\subsubsection{Simulation, Experimentation, and Result Generation}

LLM agents can conduct wireless experiments across environments with increasing physical realism, from numerical simulation and site-specific propagation modeling to potential real-world radio frequency (RF) measurements. Through programmatic tool access, they can invoke MATLAB or Python runtimes for numerical studies, use the NVIDIA Sionna library for link- and system-level simulations, and construct site-specific radio-propagation scenarios with Sionna RT~\cite{hoydis2022sionna,hoydis2023sionnart}. With appropriate hardware interfaces, human authorization, and safety controls, the framework could be extended to enable agents to interact with USRP-based software-defined radios through UHD, GNU Radio, or equivalent interfaces. Using the selected environment, the agents convert the system assumptions, algorithm specification, and evaluation criteria into executable experiment configurations. They implement the proposed method and comparison methods, explore the relevant operating regimes through parameter sweeps or repeated measurements, and organize the raw outputs into performance curves and tables. Execution checks identify inconsistencies between the implementation, experiment configuration, and mathematical formulation, prompting targeted revision when necessary~\cite{guo2026wara}.

\subsubsection{Scientific Insight Synthesis and Manuscript Preparation}

Validated analytical and experimental results are jointly analyzed to identify reproducible performance trends, governing factors, and deviations from theoretical predictions. LLM agents conduct cross-scenario comparisons and sensitivity analysis to determine the dependence of the observed behavior on operating conditions and modeling assumptions. This synthesis establishes the domain of validity of the original hypothesis, distinguishes findings that persist across operating regimes from configuration-dependent effects, and may reveal previously unrecognized tradeoffs or operating regions. LLM agents then use LaTeX to present the validated analysis and findings in a clear manuscript draft \cite{lu2026aiscientist,liu2026autoresearchclaw}.
\vspace{-3mm}

\subsection{Evidence-Centered Orchestration and Configurable Autonomy}

To ensure technical validity, the five wireless research stages are connected through contract-based validation gates. At the beginning of the whole process, a workflow-level contract defines the research scope, approved tools and datasets, evidence requirements, stopping criteria, and human-review policy. For each stage, a stage-specific contract further specifies its inputs, outputs, assumptions, and acceptance criteria.

At each gate, the responsible agent produces machine-readable diagnostics and a human-readable review report. Dedicated validation agents can use the diagnostics to assess contract compliance, consistency with upstream artifacts, and evidential support, while human researchers can inspect the artifacts and review report. Depending on the configured policy, validation agents, human researchers, or both can determine whether the artifacts pass or require revision. Accepted artifacts are versioned and transferred to the next stage, whereas failed checks trigger targeted revisions. Throughout the workflow, a shared research record preserves all artifacts, dependencies, validation results, and the complete chain of evidence  \cite{zhong2026harness,guo2026wara}.

The framework supports configurable autonomy levels by distributing research tasks and approval decisions between LLM agents and human researchers. In the agent-assisted mode, researchers lead selected stages with support from agents. In the gate-supervised mode, agents execute each stage, while researchers review the reports and approve designated handoffs~\cite{liu2026autoresearchclaw}. In the end-to-end mode, agents coordinate the complete workflow under a predefined research contract, with researchers providing the initial wireless topic and conducting the final assessment \cite{guo2026wara}. An autonomy mode can be selected independently for each stage, allowing human involvement through direct execution, intermediate approval, or final review.

\begin{table}[t]
	\centering
	\caption{Top five candidate UAV-ISAC research ideas generated and evaluated by the proposed AutoResearch framework.}
	\label{tab:top5-uav-isac-topics}
	\footnotesize
	\setlength{\tabcolsep}{3pt}
	\renewcommand{\arraystretch}{1.12}
	
	\begin{tabularx}{\columnwidth}{
			@{}c>{\raggedright\arraybackslash}Xc@{}
		}
		\toprule
		
		\textbf{Rank} &
		\multicolumn{1}{c}{\textbf{Research Topic}} &
		\textbf{Score} \\
		
		\midrule
		
		1 &
		\textbf{Near-Field ISAC for Six-Dimensional UAV Pose Estimation} &
		9.86 \\
		
		2 &
		High-Speed UAV Sensing With 5G NR Signals: Ambiguity Analysis and Pilot Design &
		9.82 \\
		
		3 &
		Quickest UAV Detection With Adaptive ISAC Beam Sweeping &
		9.79 \\
		
		4 &
		Recovering UAV Micro-Doppler From Punctured OFDM Signals &
		9.76 \\
		
		5 &
		Noncoherent Cell-Free ISAC for UAV Detection &
		9.73 \\
		
		\bottomrule
	\end{tabularx}
	\vspace{-5mm}
\end{table}
\vspace{-3mm}
\section{Case Study}
\label{sec:case_near_field_uav_pose}

In this section, we present a case study on “ISAC for UAVs” to illustrate how wireless AutoResearch operates. The topic builds on the authors’ extensive research experience in both ISAC and UAV communications, two key research areas for 6G and beyond. The case study demonstrates how LLM agents support five interconnected research stages, from evidence-grounded research direction selection to traceable scientific reporting. In this case study, we adopt an agent-assisted mode, in which human researchers and LLM-based validation agents jointly examine the artifacts produced at each stage. The human researchers also provide feedback on these artifacts, guiding the LLM agents in refining the results.

\begin{figure*}[!t]
	\centering
	\includegraphics[width=0.85\linewidth]{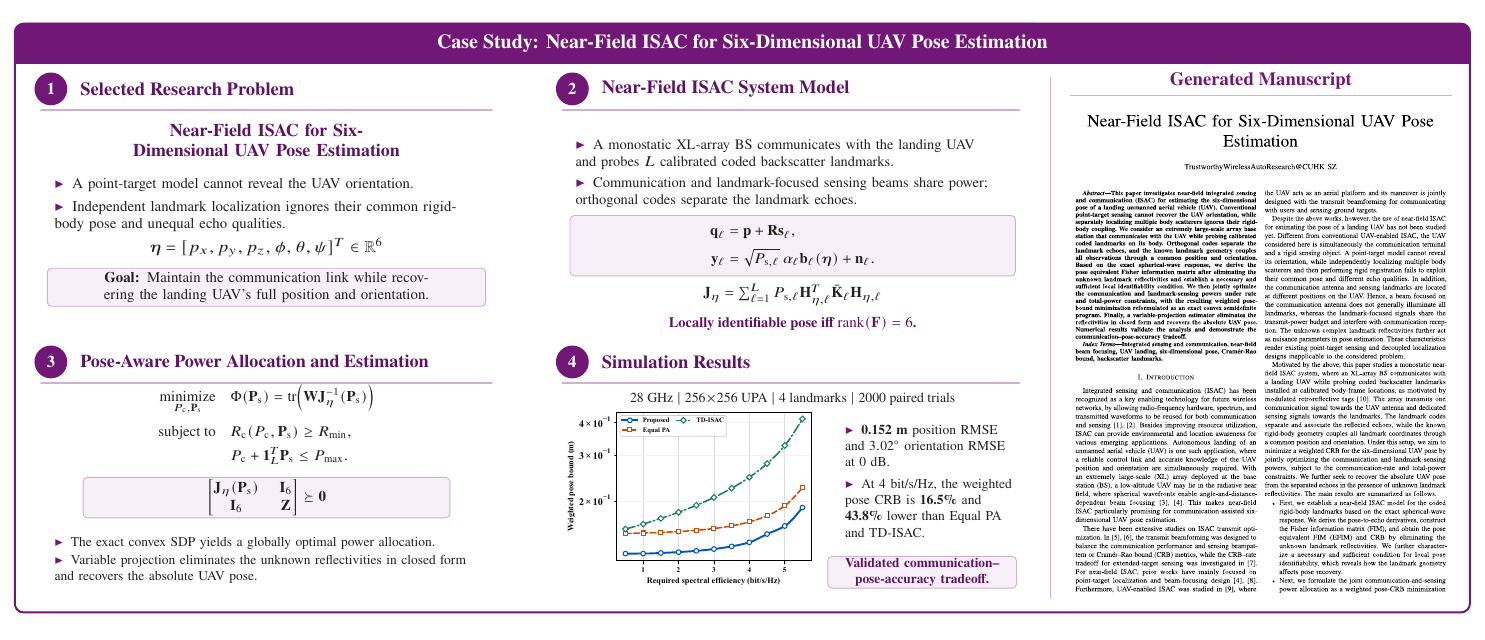}
	\vspace{-5mm}
	\caption{Case study of near-field ISAC for six-dimensional UAV pose estimation.}
	\vspace{-5mm}
	\label{fig:near-field-isac-case-study}
\end{figure*}
\vspace{-3mm}
\subsection{Candidate Idea Generation and Selection}

We begin with the broad wireless research topic ``ISAC for UAVs.'' The LLM agents then search the relevant literature and retrieve 434 papers spanning 2014--2026. Through a comparative analysis of these studies, they generate 70 candidate research ideas and evaluate each candidate across eight dimensions: problem significance, novelty defensibility, physical-layer depth, model completeness, practical relevance, literature grounding, scope discipline, and execution feasibility. The 5 highest-ranked ideas and their corresponding scores are presented in Table~\ref{tab:top5-uav-isac-topics}.

We examine the top-ranked idea, “Near-Field ISAC for Six-Dimensional UAV Pose Estimation,” in greater detail. This idea emerges from a structured comparison across several dimensions, including the UAV’s role, propagation regime, target representation, estimated parameters, sensing observations, and communication requirements. Cross-document analysis differentiates the proposed problem from existing studies that primarily model the UAV as either an aerial platform or a point target. It further reveals that independently localizing multiple body scatterers fails to fully exploit their common rigid-body geometry and heterogeneous echo qualities.

Based on this evidence, the agents formulate a focused research question: How can a large antenna array maintain a reliable communication link with a UAV while exploiting near-field observations from multiple body-fixed landmarks to jointly estimate its position and orientation? Following human review and approval of its novelty and research potential, this candidate idea is selected for further investigation.

\vspace{-3mm}
\subsection{System Modeling and Performance Metrics}

Once the direction is selected, LLM agents translate it into a research contract specifying the network roles, landmark geometry, propagation model, available observations, nuisance parameters, resource constraints, and evaluation criteria. The resulting system uses a monostatic BS that transmits a communication signal toward the UAV antenna and dedicated sensing signals toward calibrated coded backscatter landmarks. Orthogonal probing signals and landmark codes separate and associate the reflected echoes, while the known rigid-body geometry couples all landmarks through one common UAV pose.

Using the exact spherical-wave response, the agents support the derivation of the equivalent pose information and the corresponding Cram\'er--Rao bound (CRB) after eliminating the unknown landmark reflectivities. They also establish how the landmark geometry and the available sensing information determine local pose identifiability. Independent validation agents check the coordinate transformations, signal dimensions, nuisance-parameter treatment, analytical derivatives, and identifiability conditions before the model is approved for algorithm development.
\vspace{-3mm}
\subsection{Problem Formulation and Optimization Algorithm Design}

With the physical and analytical model established, LLM agents formulate how the shared transmit power should be divided between communication and landmark sensing. With the communication and sensing beams fixed, they construct a weighted pose-bound minimization problem subject to the communication-rate requirement and transmit-power constraints. By exploiting the structure of the communication constraint and pose-information matrix, the agents reformulate the power-allocation problem as an equivalent convex semidefinite program that can be solved globally within the fixed-beam design.

The agents further develop a variable-projection estimator for absolute UAV pose recovery. The estimator eliminates the unknown landmark reflectivities in closed form and iteratively refines the common UAV position and orientation from the code-separated echoes. For comparison, the agents implement equal power allocation, time-division ISAC, and independent landmark localization followed by rigid registration under matched resource conditions. Validation agents examine the optimization reformulation, constraint satisfaction, and consistency between the estimator and the analytical pose model.
\vspace{-3mm}
\subsection{Simulation Execution and Performance Evaluation}

The analytical and algorithmic artifacts are then converted by LLM agents into an executable numerical workflow. The workflow evaluates the communication--pose-accuracy tradeoff, estimator behavior under different sensing conditions, unequal landmark visibility, and pose recovery over a simulated UAV descent. The proposed methods and comparison schemes use consistent array configurations, resource budgets, communication requirements, and observation conditions, while the experiment configurations, implementation code, and generated results are retained as versioned artifacts.

The evaluation produces three consistent observations. Pose-aware power allocation improves upon equal allocation and time-division operation because it accounts for the different pose-information contributions of the landmarks. The variable-projection estimator approaches the analytical pose bound and improves upon independent landmark localization by exploiting the common rigid-body pose and landmark-dependent echo quality. During the simulated descent, the position and orientation estimates generally become more accurate as the UAV approaches the BS, where stronger near-field wavefront curvature provides more informative spatial signatures.

\subsection{Synthesis of Traceable Scientific Claims}

In the final stage, LLM agents connect the numerical observations to their physical and mathematical mechanisms. They relate pose recovery to spherical-wave curvature and rigid-body coupling, the resource-allocation gain to differences in landmark visibility and rotational leverage, and the landing behavior to the distance-dependent strength of near-field signatures. A claim--evidence ledger links each manuscript claim to its supporting literature evidence, research contract, analytical result, algorithm version, experiment configuration, and numerical output. Independent validation agents review these connections, while human researchers approve the final interpretations and publication claims. The case study therefore demonstrates how LLM agents coordinate the complete progression from research direction selection to evidence-grounded scientific reporting.
\vspace{-3mm}
\subsection{Manuscript Evaluation}
To evaluate the effectiveness of the proposed framework in supporting end-to-end wireless research, we conduct a manuscript-level assessment of the resulting case study. Using the review agent introduced in WARA~\cite{guo2026wara}, we compare the case-study manuscript with five published IEEE ICC/GLOBECOM papers and five published IEEE letters under seven different backbone LLMs. As shown in Table~\ref{tab:research-validity-model-scores}, the case study receives the highest score from GPT-5.5, Qwen3.8 Max, GLM-5.2, MiniMax M3, and Kimi K2.6, and ties with Letter~2 for the highest score under DeepSeek V4 Pro. GPT-5.6 Sol is the only exception, assigning 71 to Conference~2 and 66 to the case study. Overall, six of the seven backbone models rank the case study first or jointly first, indicating that the research outcome developed through the proposed framework compares favorably with closely related peer-reviewed work across different LLM-based evaluations.

\begin{table}[t]
	\centering
	\caption{Research validity scores across seven backbone LLMs.}
	\label{tab:research-validity-model-scores}
	\footnotesize
	\setlength{\tabcolsep}{1.4pt}
	\renewcommand{\arraystretch}{1}
	\begin{tabular}{@{}lccccccc@{}}
		\toprule
		
		\textbf{Paper} &
		\shortstack[c]{\textbf{GPT-5.5}} &
		\shortstack[c]{\textbf{GPT-5.6}} &
		\shortstack[c]{\textbf{DeepSeek}} &
		\shortstack[c]{\textbf{Qwen3.8}} &
		\shortstack[c]{\textbf{GLM-5.2}} &
		\shortstack[c]{\textbf{MiniMax}} &
		\shortstack[c]{\textbf{Kimi}} \\
		
		\midrule
		
		Conf.~1 & 71 & 67 & 72 & 88 & 73 & 74 & 72 \\
		Conf.~2 & 78 & \textbf{71} & 72 & 91 & 73 & 75 & 72 \\
		Conf.~3 & 55 & 54 & 52 & 87 & 68 & 67 & 55 \\
		Conf.~4 & 62 & 56 & 55 & 85 & 64 & 64 & 52 \\
		Conf.~5 & 40 & 45 & 52 & 81 & 62 & 48 & 58 \\
		
		\midrule
		
		Lett.~1 & 74 & 61 & 62 & 82 & 68 & 62 & 58 \\
		Lett.~2 & 75 & 69 & \textbf{78} & 92 & 75 & 74 & 71 \\
		Lett.~3 & 65 & 52 & 62 & 85 & 68 & 67 & 54 \\
		Lett.~4 & 72 & 66 & 68 & 88 & 68 & 70 & 68 \\
		Lett.~5 & 61 & 52 & 48 & 73 & 59 & 59 & 38 \\
		
		\midrule
		
		\shortstack[l]{\textbf{Case}\\\textbf{Study}} &
		\textbf{79} &
		66 &
		\textbf{78} &
		\textbf{96} &
		\textbf{85} &
		\textbf{82} &
		\textbf{78} \\
		
		\bottomrule
	\end{tabular}
		\vspace{-5mm}
\end{table}

\section{Concluding Remarks}

This paper has presented an evidence-centered framework for wireless AutoResearch, in which LLM agents support the full research lifecycle, from literature review and idea generation to modeling, experimentation, insight synthesis, and manuscript preparation. Trustworthiness is ensured through research contracts, versioned artifacts, independent validation, and configurable human oversight. The case study on near-field ISAC for six-dimensional UAV pose estimation demonstrates how the framework transforms a literature gap into a well-defined research problem, executable evidence, and appropriately bounded claims. It also highlights the importance of maintaining consistency across assumptions, information structures, algorithm designs, optimization variables, baselines, numerical results, and scientific interpretations.

It is worth noting that the wireless AutoResearch framework is not limited to developing agentic solutions. Its outputs may include transceivers, waveforms, estimators, resource-allocation strategies, protocols, and information-theoretic bounds, which can be implemented through conventional signal processing, agentic controllers, or hybrid architectures. As a result, wireless AutoResearch expands and validates the wireless design space without prescribing a particular implementation paradigm.

It is also worth noting that compared with WARA’s minimal-human-intervention setting \cite{guo2026wara}, the case study demonstrates that appropriate expert involvement can yield stronger results, underscoring the value of human–AI collaboration. While LLM agents can accelerate information retrieval, implementation,
validation, and iterative refinement, human researchers should retain
responsibility for scientific significance, modeling choices, safety, interpretation, and publication decisions. With stronger provenance, formal validation, reproducible benchmarks, and controlled interfaces to simulation and experimental platforms, wireless AutoResearch can improve research efficiency, reproducibility, and transparency while preserving scientific accountability.

\bibliographystyle{IEEEtran}
\bibliography{references}

\end{document}